# Object-oriented modelling with unified modelling language 2.0 for simple software application based on agile methodology

H.L.H.S. Warnars*

*Department of Computing and Mathematics, Manchester Metropolitan University, John Dalton Building, Manchester, M15 6BH, United Kingdom*

Unified modelling language (UML) 2.0 introduced in 2002 has been developing and influencing object-oriented software engineering and has become a standard and reference for information system analysis and design modelling. There are many concepts and theories to model the information system or software application with UML 2.0, which can make ambiguities and inconsistencies for a novice to learn to how to model the system with UML especially with UML 2.0. This article will discuss how to model the simple software application by using some of the diagrams of UML 2.0 and not by using the whole diagrams as suggested by agile methodology. Agile methodology is considered as convenient for novices because it can deliver the information technology environment to the end-user quickly and adaptively with minimal documentation. It also has the ability to deliver best performance software application according to the customer's needs. Agile methodology will make simple model with simple documentation, simple team and simple tools.

**Keywords:** unified modelling language; UML 2.0; object oriented; agile methodology; software engineering

## 1. Introduction

The unified modelling language (UML) as analysis and design tool for object-oriented paradigm was founded by the trio Grady Booch, James Rumbaugh and Ivar Jacobson. Grady Booch with his object-oriented design concept, James Rumbaugh with his object modelling technique and Ivar Jacobson with his object-oriented software engineering have strengthened the UML's ability as a powerful tool for object-oriented analysis and design. Rumbaugh in October 1995 together with Booch yielded the unified method 0.8. In June 1996, Jacobson joined them and yielded UML 0.9. Finally, UML 2.0 was introduced in 2002 with 13 diagrams (Booch 1999, Larman 2005).

Nowadays, software programming uses structured paradigm rather than object-oriented paradigm, and hence the world of information system development also transforms from structured paradigm to object-oriented paradigm. UML has been known as a modelling language for object-oriented paradigm and has been used as an official standard for data modelling, business modelling and real-time development. Those who want to model the system with current object-oriented paradigm must consider UML as an approved modelling language. As a modelling language, UML consists of notation and set of rules like syntax, semantics and pragmatics, which must be understandable for those who want to learn this modelling language (Eriksson 2004). UML is not a programming language but a modelling language which can be used to model the information system with some drawing design by the diagram's rule and where the diagram will be used as a communication tool within a team.

To draw or model the information system as a communication tool within a team or with the customer, the model will need the methodology or method that can be used as an explicit way to think and act (Eriksson 2004). For some writers, methodology is different from method (Bennett *et al.* 2002). A methodology or software engineering method consists of process, standard vocabulary, set of rules and guidelines which tell what to do, how to do, when to do and why it is done (Eriksson 2004, Sommerville 2004). Agile methodology as an evolutionary development will more effectively meet the needs of the customer – quickly, adaptively and focus on the software itself. In the beginning, agile methodology was suitable for small- or medium-sized business and personal computer products (Sommerville 2004), but soon this has been implemented in large-sized business as well (Holmstrom *et al.* 2006, Ambler 2008f).

Documentation and team effectiveness will help in building the best performance software application as per the customer's requirement. Agile methodology makes simple models using simple tools (Ambler 2002).

## 2. Problem definition

There are a lot of sources like the internet, books, papers or journals which can be used as references to design the object-oriented information system with UML, especially with UML 2.0. Each source can use different UML versions and there is even some overlap in the use of UML versions in UML diagrams. In one diagram, they may use UML version 1.x and in the other diagram they may use UML version 2.0. Moreover, for some samples or methodologies like agile modelling has combined between UML as an object-oriented modelling with data flow diagram (DFD) as structured modelling.

Furthermore, UML diagrams have been criticised for deficiencies such as semantic inconsistencies, inadequacy of notation and ambiguities of diagrams and construction (Ambler 2002, Siau and Loo 2006). There are ambiguities in diagrams like sequence diagram, communication diagram, activity diagram and state machine diagram. We can design our interaction diagram with sequence diagram or communication diagram. In the same way if we want to design behaviour diagram we can design with activity diagram or state machine diagram. Figure 1 shows some UML complaints.

For agile modelling UML which encouraged to use for applying UML (Larman 2005) is not sufficient for the development of business software and more complex (Ambler 2002, Ambler 2008b). The question 'What's missing from the UML?' seems to judge whether the UML is strange, too little in it and too much in it (Ambler 2002). If something is wrong with it, why should there be a statement like that? Is there any competition with Rational Rose, or with the UML's founder? Another issue is that there is a lack of computer assisted software engineering (CASE) tools which can only do the static models. Reverse engineering can generate class diagram from source code but it cannot generate interaction diagram like sequence diagram. On the other hand, in forward engineering, the code programming can generate from class diagram but not the method body code from interaction diagram like sequence diagram (Larman 2005). These questions become more interesting to dig in more.

## 3. Object-oriented as it is

As a UML learner, one should never get involved with object-oriented mailing list and should have some experience to implement UML in object-oriented analysis and design; the following represents my experience in learning UML 2.0:

(1) There is a lack of UML 2.0 resources specifically for Indonesian books.
(2) There are not many example cases to analyse and design with UML specifically with fully UML 2.0 and also for Indonesian books. If be present then will have different ones and need extra time to conquer the differentiation.
(3) Learning will be easier if we have understood object-oriented concepts like class, object, attribute or property or data, method or function or behaviour, inheritance, multiple inheritance, encapsulation, polymorphism, etc.
(4) Learning will be easier if we have understood and implemented some object-oriented programming (Siau and Loo 2006) and conversant to using object-oriented concepts like class, object and inheritance by programming implementation. It is not only important to just know how to use object-oriented program but one should also know how to implement the object-oriented concept in programming. One should also know how to implement class, object, inheritance, encapsulation, polymorphism in programming, as shown in Figure 2. I think it is very important that one must know how to use object-oriented programming with a truly object-oriented concept. Also, for each object-oriented programming there are some differentiations for using it, e.g. with C++, Java, Visual Basic, etc. For example, in Java we can do multiple inheritance but in C++ we cannot do that. In Visual Basic, we can implement without being fully object oriented. Visual Basic is different from C++ or Java. In view of all the above-mentioned requirements, one should possess the ability to use and implement object orientation and also should be aware that there are different concepts to implement for each object-oriented programming.

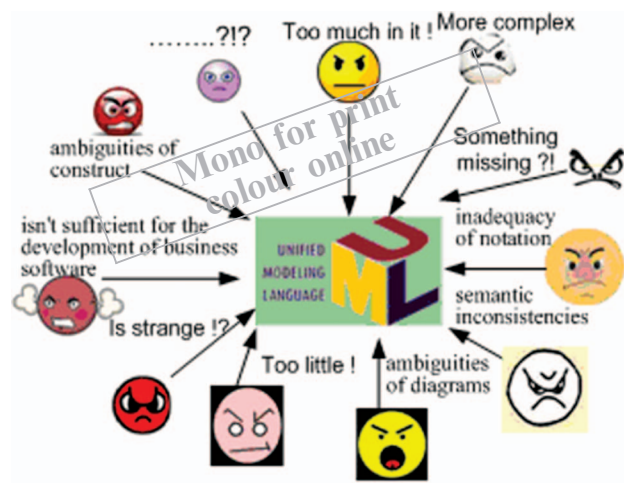

Figure 1.    Complaints about UML.

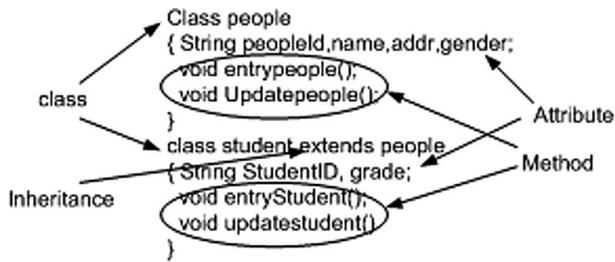

Figure 2. Object-oriented concepts.

Beyond the UML versus non-UML paradigm, the acknowledgement from industry leaders like IBM, HP, Oracle, Microsoft, etc. and Object Management Group (OMG) standardisation for UML became the most interesting parameter for the question, 'Why are people not positive about UML?' I was also having negative opinion when I started learning UML for the first time, because I have always used structured paradigm. Hence, I tried to incorporate structured paradigm in my object-oriented concept and sometimes tried to mix between structured tools paradigm like Entity Relationship Diagram, DFD, Flowchart and object-oriented paradigm with UML. One may ask why some people had a negative view about UML. The answer could be that they did not know how to use UML diagram or they must be new to UML.

The acknowledgement from industry leaders and OMG about UML is one of the parameters for me and my organisation in choosing UML as an approved object-oriented modelling language. Initially, it would be difficult but gradually we can change our mindset from structured paradigm to object-oriented paradigm. I am not a UML seller but I think as a human I have always liked the introduction of a new technology.

My knowledge about agile methodology has helped me to learn and use UML in my modelling documentation. We must experiment with our model and should not strictly follow a particular process to model the system because different people might have used the stereotype in different manners (Ambler 2002). One of Ambler's arguments is that UML does not include user interface diagram, but the question is why he has not pictured out his user interface diagram with UML diagram like sequence diagram or activity diagram? Like his agile modelling statement 'Small is better', design simple model with simple tools but not too simple as sometime become complex tools (Ambler 2002, Koch 2008). It is contradictory that why we are not using only small tool in one language modelling and do not cross to other language modelling or paradigm.

To implement agile modelling as a small and simple concept for novice, it will be better (Meso and Jain 2006, Vinekar *et al.* 2006) if we can explain and teach them with simple and small tools rather than using many diagrams from many paradigms and modelling languages. Of course, using many diagrams from many concepts or crossing paradigms will make ambiguities and inconsistencies for novice to learn how to model the system. In my opinion, UML is the powerful and the right modelling language. Even though UML has many diagrams, we just use the diagram that we need as simple as we need as agile methodology's theory.

Mixing notation between structured and object-oriented paradigm will make our modelling language complicated. If we implement our system with object-oriented programming then we must quit the structured paradigm and transform to object-oriented paradigm. On the other hand, object-oriented models are conveniently implemented with object-oriented programming and the non-usage of object-oriented programming is not recommended (Eriksson 2004). If we still use structured paradigm could be we still put structured paradigm in our head and cannot leave it and always think structured when we think about object-oriented paradigm.

## 4. Consistency

Now, we will try to create a sample design with UML 2.0. Perhaps this will start a new discussion, but I think every person or institution has a right to use all the diagrams in UML 2.0 as they need as long as it will help them and make it easier for their business process in line with the purpose of the analysis and design information system to satisfy the users. In this sample design we do not use all the diagrams but just use case diagram, class diagram, sequence diagram and package diagram as the requirement for agile methodology to deliver the system in a simple manner.

When performing the analysis and design, we should care about the consistency where there will be a relation between the design and implementation. We must design something that we have implemented or we must implement something that we have designed.

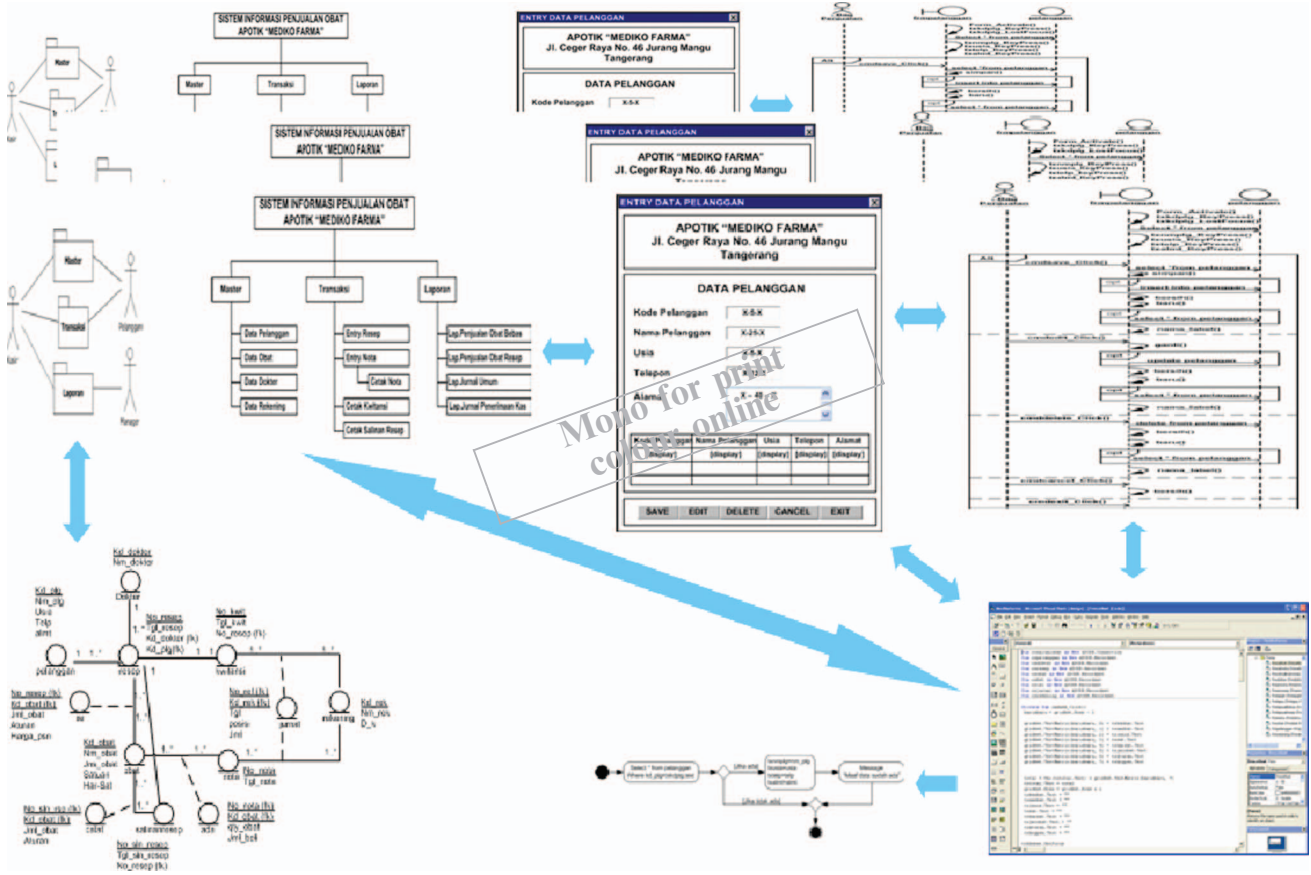

Figure 3. Redline between design and implementation as consistency.

There will be a red line between design and implementation (Booch 1999). This consistency will be run as the need of user requirement and every step will influence other steps because changing one step will change other steps as well; changing the onward step will influence the backward step and vice versa. Figure 3 shows how consistency can be done as relation between design and implementation.

The same happens in other fields of science, for example accountancy. We can trace net income or net loss in income, with the help of the transaction entries made in ledger or journal. Figure 4 shows the accounting cycle as consistency from each of the steps in the accounting cycle.

The same happens in architecture as well; the job of the builder is just to build something which has been drawn on blueprint. If the builder could not accomplish the job according to the blueprint, it will mean that the builder does not have the ability to read the design or blueprint. Figure 5 shows an example of a house blueprint.

Building the information can be done with top-down approach or bottom-up approach or a combination of both.

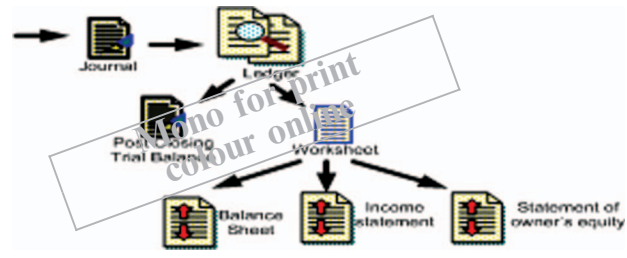

Figure 4. Accounting cycle.

## 5. Software modelling lifecycle

To model a simple software application, we need some steps as guidance. One can refer to agile methodology to make the steps for modelling the software application as simple as possible, easy to learn and for best performance implementation to fulfil user's requirements. In the beginning, agile methodology was suitable for small- or medium-sized business and personal computer product (Sommerville 2004), but later it has been implemented in large-sized business as well.

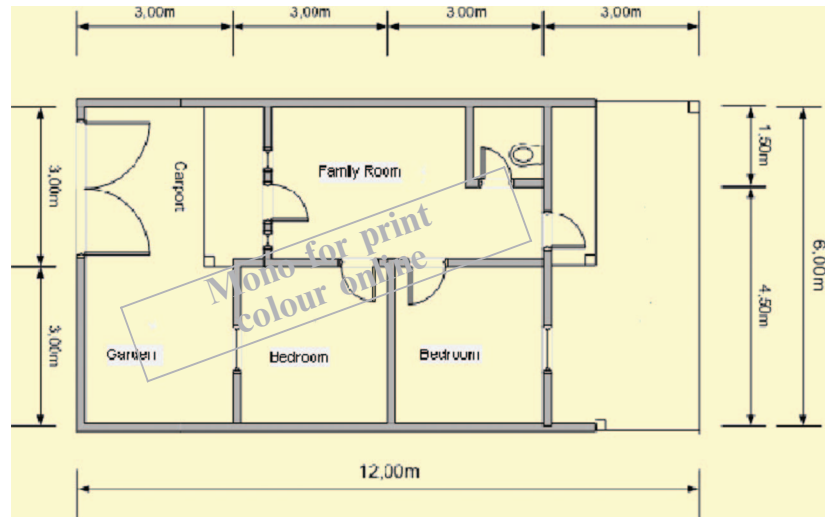

Figure 5. Blueprint of a house.

As agile methodology theory about using simple tools for developing software application we formulate software modelling lifecycle as shown in Figure 6. This software modelling will be sufficient for all kinds of business level/area according to my experiences with the implementation of object-oriented analysis and design. Each software modelling lifecycle can be chosen by every organisation or person to match their business process and thereby help in the development of the business. Sometimes an organisation or a person will explore some software modelling lifecycle without the need to use the whole steps and even they can add some other methods for improvement. There are some advantages and disadvantages in all the software modelling lifecycles.

The steps for developing software application in Figure 7 will be divided as follows:

(1) User requirement
(2) Data model
(3) Business process model
(4) (a) Prototyping; (b) Class interaction.

In the fourth step there will be simultaneous activities. This step can be started with prototyping or class interaction activities but each of them will complement each other. For each step there will be technique, tools and diagrams which will be arranged chronologically.

### 5.1. User requirement

User requirement bridges our knowledge with the user's needs. We would never know the needs, expectation, dream of the user if we do not analyse the user requirement. User requirement will be done in order to satisfy the user. For user requirement we do not need UML diagram, but common user requirement techniques like interview, observation, documents, survey, Joint Application Development (JAD) and other techniques are sufficient.

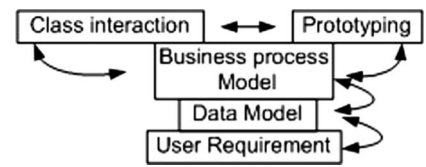

Figure 6. Software modelling lifecycle.

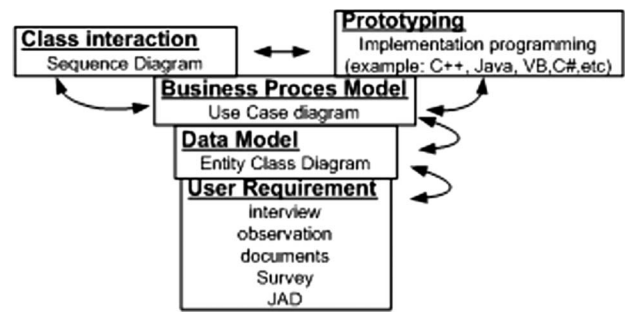

Figure 7. Software modelling lifecycle with technique, tools and diagram.

There are five user requirements which can be used by all, but the first three must be done for novice. Interview, observation and documents must be carried out because when we perform user requirement then we must interview the user in a non-formal manner. While interviewing the user should be able to observe how the business process is run. Meanwhile when doing the observation we will collect all the documents, both input and output documents, which flow on the as-is business process. If we miss some documents then we must search for those documents.

For the novice, the first three techniques like interview, observation and document verification must be done to know about the knowledge and experience of the user. For the non-novice, these techniques are not needed if they have knowledge and experience for the as-is business process. For completed documentation, it is necessary that these three techniques are used in a simple manner.

Survey technique can be used for both novice and non-novice as a supplement to strengthen our model justification. Survey is not acceptable if the user does not fill in any details. Survey for the future will become a knowledge justification for the next modelling software application but cannot be different in any kind of business process, organisation, age, etc. The skills and tricks for making questions and choosing the type of questions in survey will be needed to create a comprehensible question. For an expert, the survey will become the shortcut data to model the software application.

JAD as a technique with large audience gathers staff from low level to high level management. JAD will need much cost, time and efforts. JAD will be needed as a complement by user's agreement. For medium-sized business, JAD could be a breakthrough to deliver the best information technology business process that has been approved by all level managements.

Using and choosing requirement techniques are dependent on the size of the system that we will be building. If we build a small system that does not involve many people then we can choose none of the techniques, but if we build medium or large system and involve many people then we need to use requirement techniques as a documentation and communication between people in team. We can use interview and observation techniques in non-formal situations and without documentation when getting the information from the user and in order to satisfy our user. Sometimes people always using interview and observation techniques but they have not realised if they used it as they did not do the documentation when doing interview or observation. Sometimes people do not realise when they use these techniques when they are talking to their user in non-formal manner. People can build the system without doing user requirement techniques, but it is better to use them. If we build the system then we must satisfy our user, produce high quality software that meets the user needs and to meet the user need, to satisfy our user then at least we have to talk with them, in non-formal manner and in effective manner. And the same as the most things, with one of agile principles is to satisfy our user not ourselves (Ambler 2002).

### 5.2. Data model

After collecting the user requirements to satisfy user, based on observation documents and interview stories, then we will model the data as database. The line arrow between user requirement and data model step, as shown in Figure 8, shows that data model step can be done only if user requirements activity has been done. Also the activity in data model will influence user requirement. When do the data model activity we can go back to user requirement step if we feel lack of documents or knowledge about business process. The second or more user requirement as a request for the lack in the first user requirement can be done.

For the sake of simplicity, sometimes the step user requirement and stepdata model can be done together as one step to model the data as database. Figure 9 shows how user requirement and data model steps become one step, different to Figure 6 where user requirement and data model steps are different steps.

There are three types of classes (Eriksson 2004, Whitten et al. 2004):

(1) Boundary class: Form design which will be used by user as communication or interface to system. Figure 10 is the symbol of boundary class.
(2) Control class: Application logic or controller where one can manage class interaction specifically to connect boundary class and entity class. Figure 11 is the symbol of control class.
(3) Entity class: Data stored as table in database. Figure 12 is the symbol of entity class.

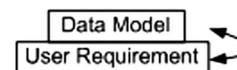

Figure 8. Line arrow between user requirement and data model.

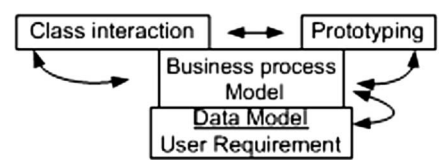

Figure 9. Simple software modelling lifecycle.

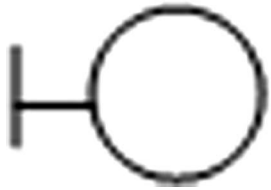

Figure 10. Boundary class.

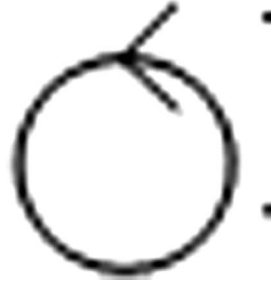

Figure 11. Control class.

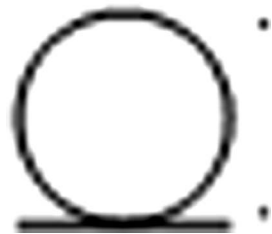

Figure 12. Entity class.

Figure 13 shows each type of class in implementation as consistency between design and implementation. Data model only uses entity class as data stored in database, boundary and control class will be used when we model class interaction with sequence diagram (Ambler 2008a).

For example, when we do user requirement we got two as-is documents for retail shop. As interview asked then has been known that every customer will be given an invoice as a bill for their transaction. Figure 14 shows an example of as-is invoice document.

After the customer has paid the bill the shopkeeper will issue the receipt as a proof of payment. Figure 15 shows an example of as-is receipt document.

Based on the two as-is documents and the interview's results then the entity class diagram will be modelled as shown in Figure 16. This entity class diagram is logical data model for each table in the database.

### 5.3. Business process model

After the step data model has been done we move on to the next step to build the business process model using case diagram. Based on entity class diagram, the business process model will be modelled. The line arrow between data model and business process model step in Figure 17 shows that business process model step can be done only if data model's activity as entity class diagram has been created. Also the activity in business process model will influence data model. We can go back to data model step if in the middle of our model we need to model new entity class or table and automatically will influence the use case diagram model.

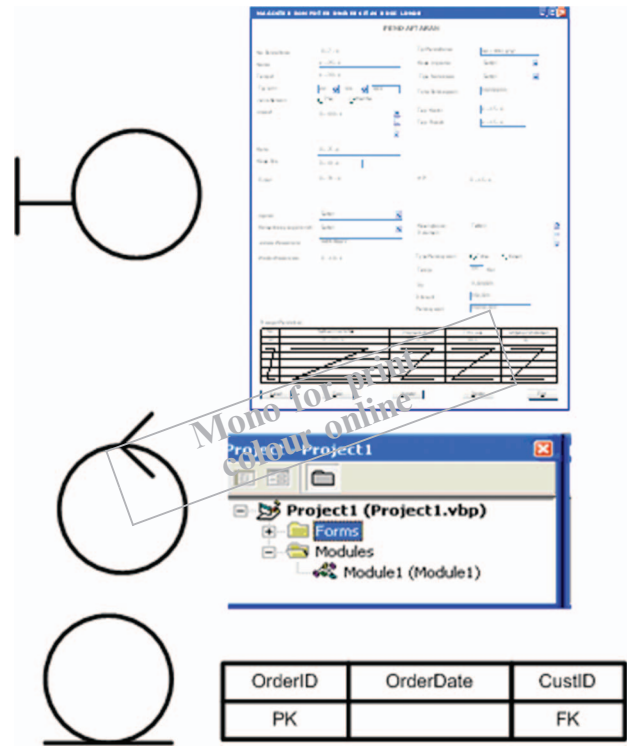

Figure 13. The three types of classes and implementations.

Before transforming our entity class diagram into business process model using case diagram, we must first divide the entity class diagram or table as follows:

(1) Master table: Master table is a table where the records must be created first before running the system. Recording at this table has to be relatively permanent.
(2) Transaction table: A table where the records must be created if there is a transaction.

Based on example entity class diagram we have two master tables which are Customer and Item tables and three transaction tables which are Invoice, Receipt and Purchase. For each master table we will create one use case on use case diagram, as shown in Figure 18. The first, use case will be named maintain customer as a process to maintain master table customer. The next

Figure 14.  As-is document – invoice.

Figure 15.  As-is document – receipt.

use case will be named maintain item as a process to maintain master table item.

For transaction we will make use case on use case diagram based on documents (Ambler 2008c) which will be created and as a result in user requirement step, we must create two documents which are invoice and receipt documents, as shown in Figure 19. The use case will be named 'create invoice' as a process to create invoice document and the other will be named 'create receipt' as a process to create receipt document at to-be system.

Each system must have some reports which can be used by high level management or owner as information to make the decisions. Many reports can be created as the user need, as shown in Figure 20, for example.

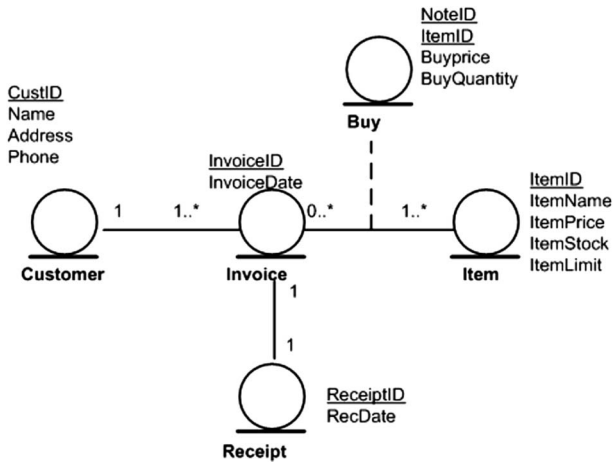

Figure 16. Entity class diagram model.

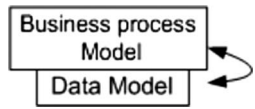

Figure 17. Line arrow between data model and business process model.

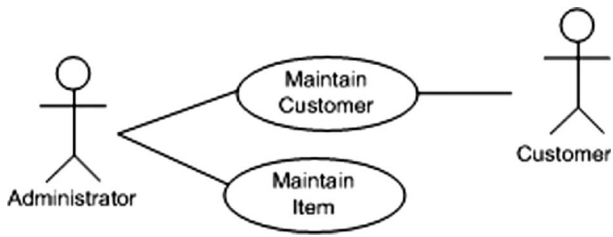

Figure 18. Use case diagram for master table.

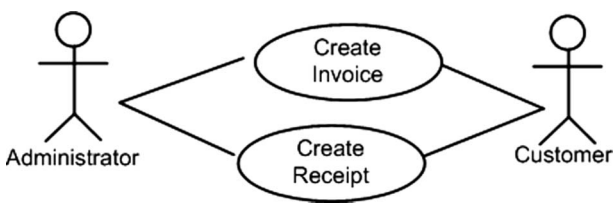

Figure 19. Use case diagram for transaction table.

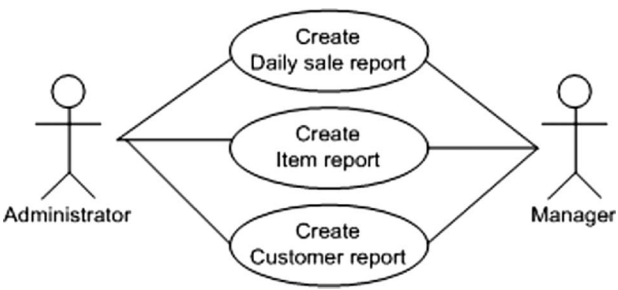

Figure 20. Use case diagram for reports.

Each view of documents or reports can be designed with form design that can be created with some software applications like MS Word or Visio. The view of documents or reports can be skipped and as a part of prototyping for simplification. Another approach can be done during the user requirement step, by observing documents and reports and the interview questions can be provided as justification. For non-novice, the view of documents or reports is a simple thing which can be written in leisure time as per the user requirement.

Notice that our use case diagram becomes simpler when we use package diagram as a neutral diagram which can be used in many UML diagrams (Ambler 2008e). Reorganising large use case diagram will be done based on the partition by master table, transaction table and reports, as shown in Figure 21. The relationship between actor and use case on use case diagram must be consistent in relation between actor and package on package diagram.

### 5.4. Prototyping and class interaction

After the business process model we can move onto the last step, i.e. prototyping and class interaction, which contains two activities which can be done simultaneously. We can start with prototyping first or class interaction with sequence diagram. Any change in each step will influence each other. If the software modelling lifecycle is made simple, these two steps can be combined as one step, as shown in Figure 22.

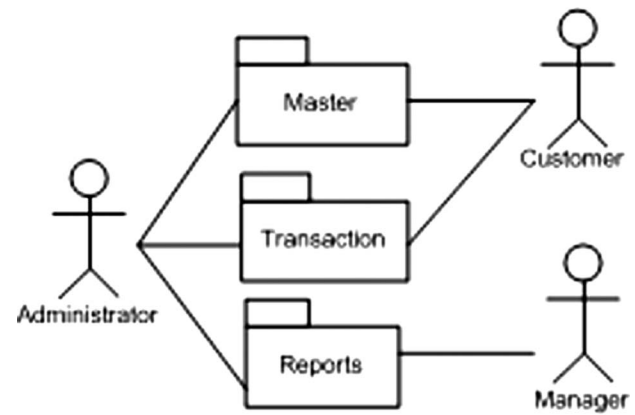

Figure 21. Package diagram.

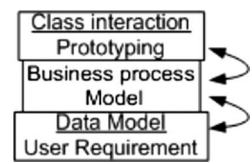

Figure 22. The simplest software modelling lifecycle.

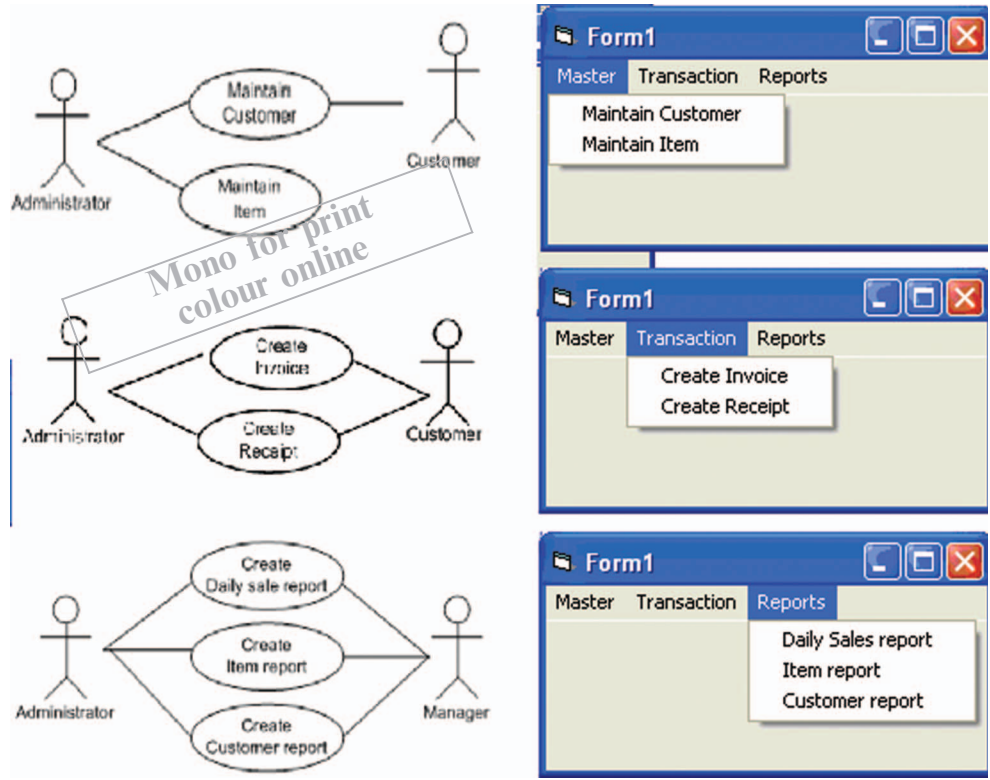

Figure 23.  Main menu prototyping.

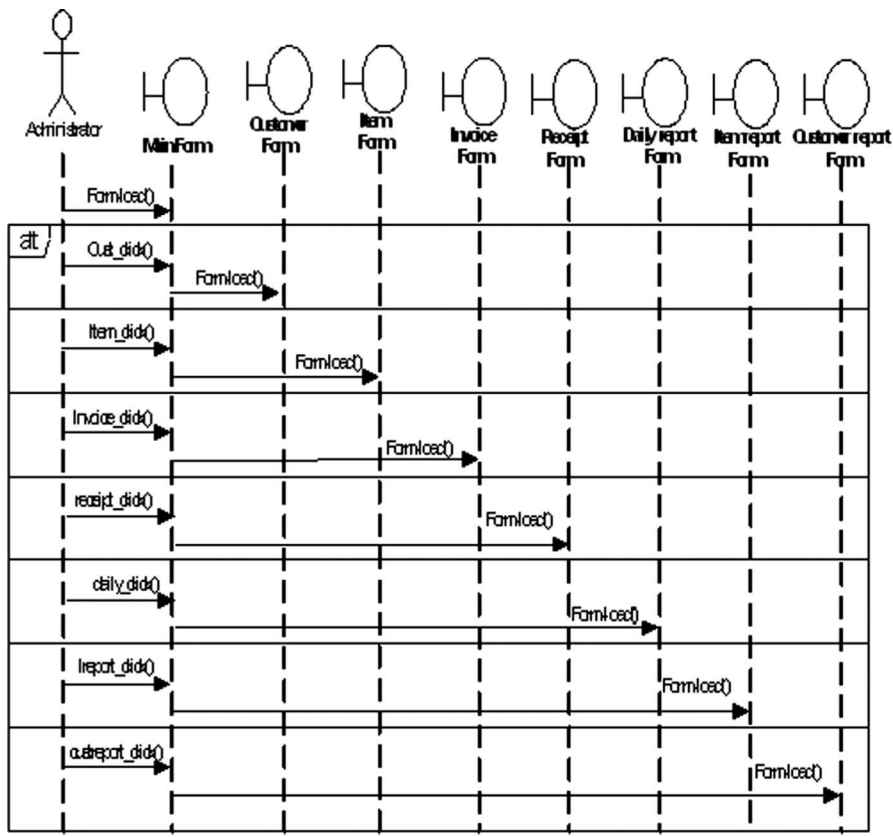

Figure 24.  Main menu sequence diagram.

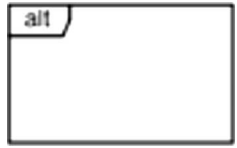

Figure 25. Alt interaction frame.

The line arrow between class interaction/prototyping and business process model step shows that class interaction/prototyping step can be done only if business process model activity with use case diagram model has been created. Also the activity in class interaction/prototyping will influence business process model. We can go back to business process model if in class interaction/prototyping need something has never been thought or predicted before. Of course, the change in business process model will also influence data model/user requirement step if necessary.

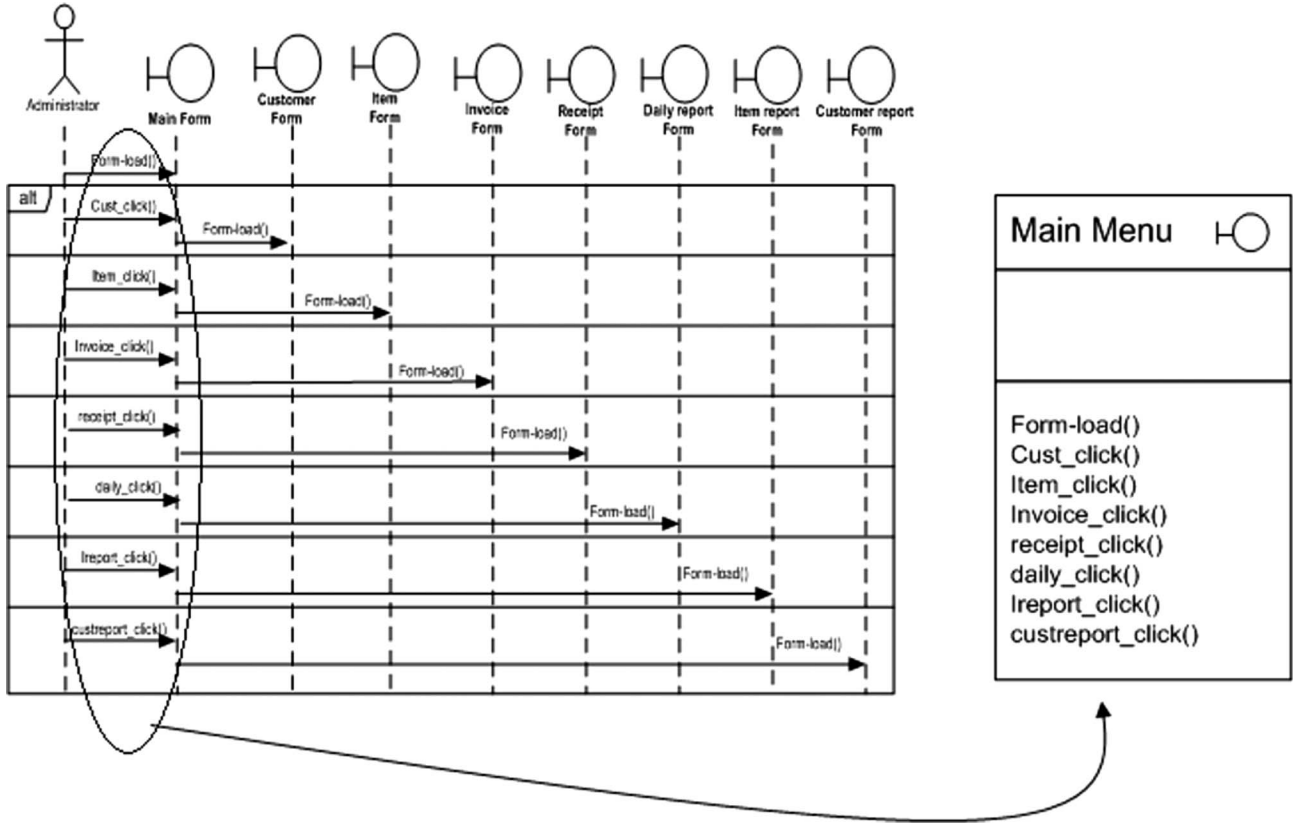

Figure 26. Place of methods.

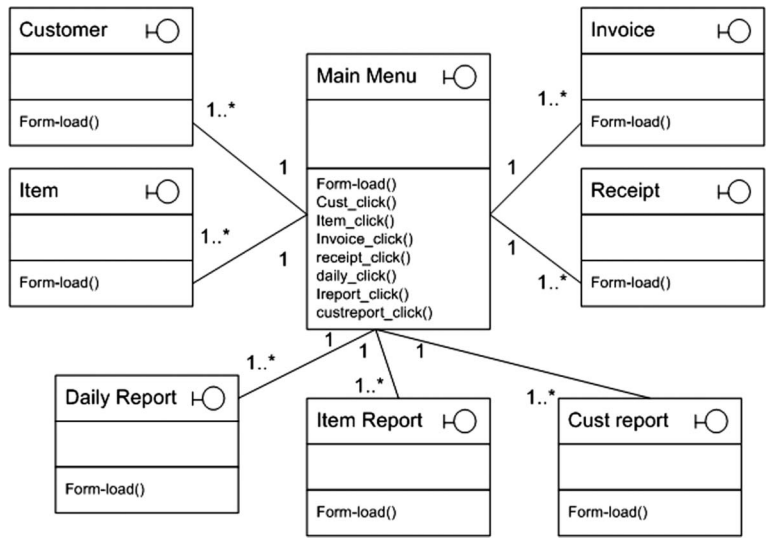

Figure 27. Boundary class diagram.

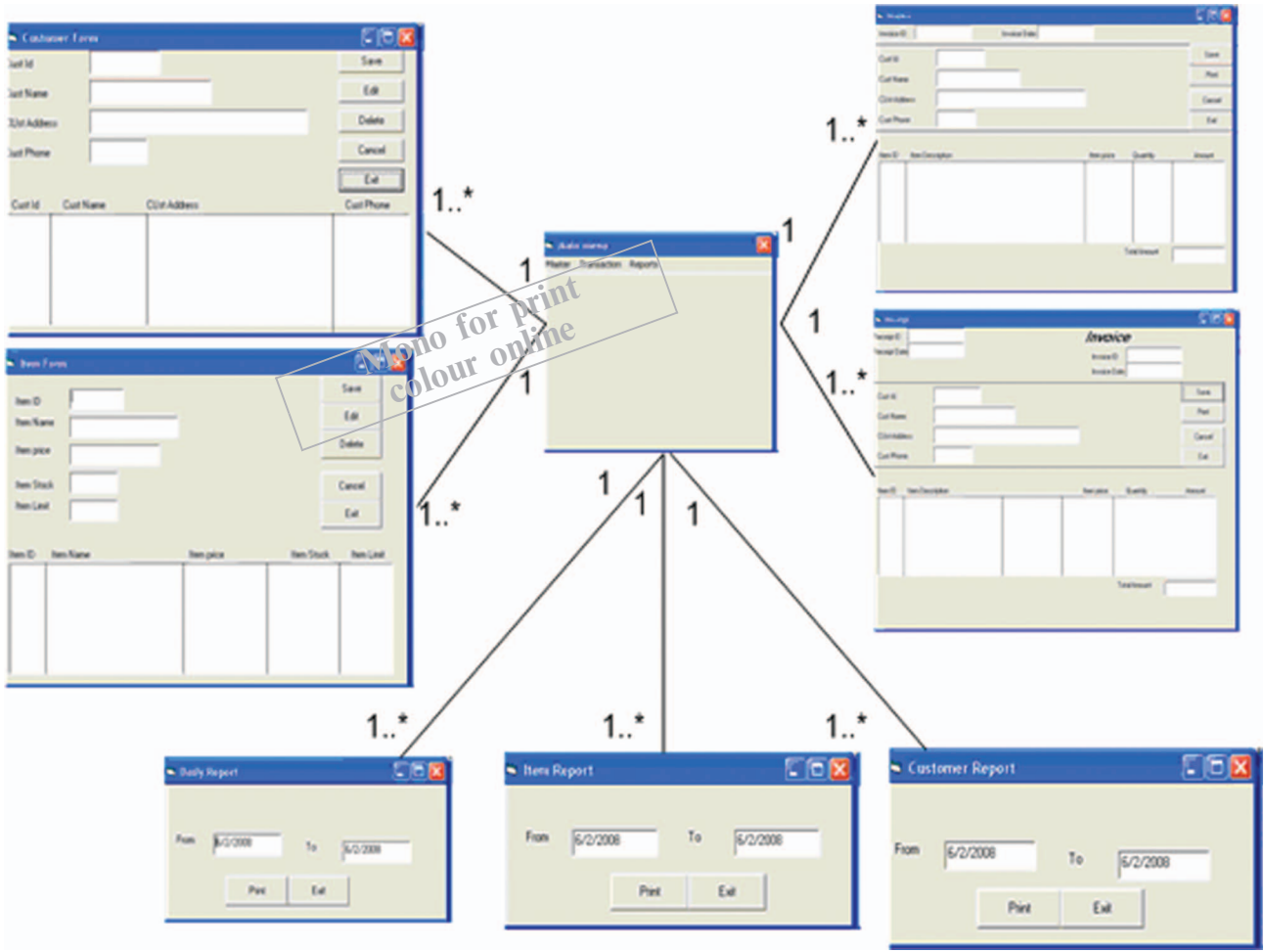

Figure 28.    Boundary class diagram VB 6.0.

For prototyping step in this article, we will use Visual Basic 6.0 as the implementation. Other program languages can also be used for the implementation but in a lesser manner as there will be a differentiation in sequence model. As we know, sequence diagram is an interaction diagram which is dynamic (Larman 2005).

Based on dividing by master table, transaction table and reports in use case diagram then the prototyping programs are shown in Figure 23.

Each use case on a use case diagram will become a menu object in the implementation of prototyping. The concept to draw sequence diagram will depend on the algorithm on prototyping implementation and the way to use software application. As with the main menu prototyping above, then based on use case diagram also we can model class interaction for main menu form with sequence diagram, as shown in Figure 24.

Each of use case on use case diagram will have a form design as a result of prototyping and a sequence diagram (Ambler 2008d).

The alt interaction frame on sequence diagram, as shown in Figure 25, is used as mutually exclusive alternatives (Larman 2005). Because we have seven use cases on use diagram and of course there are seven link menus for prototyping implementation then will be created seven alternatives in alt interaction frame as optional for actor in figure earlier. Every menu link at prototyping implementation can be clicked with the mouse or by making shortcut for keyboard access.

Form_load( ) method on picture above which will be triggered by an actor will be saved on main menu boundary class diagram. At least main menu boundary class diagram will have nine methods, as shown in Figure 26.

Each of the other boundary class diagrams will have at least form_load( ) method, as shown in Figure 27.

Boundary class diagram Visual Basic 6.0 as the prototyping implementation with Visual Basic 6.0,

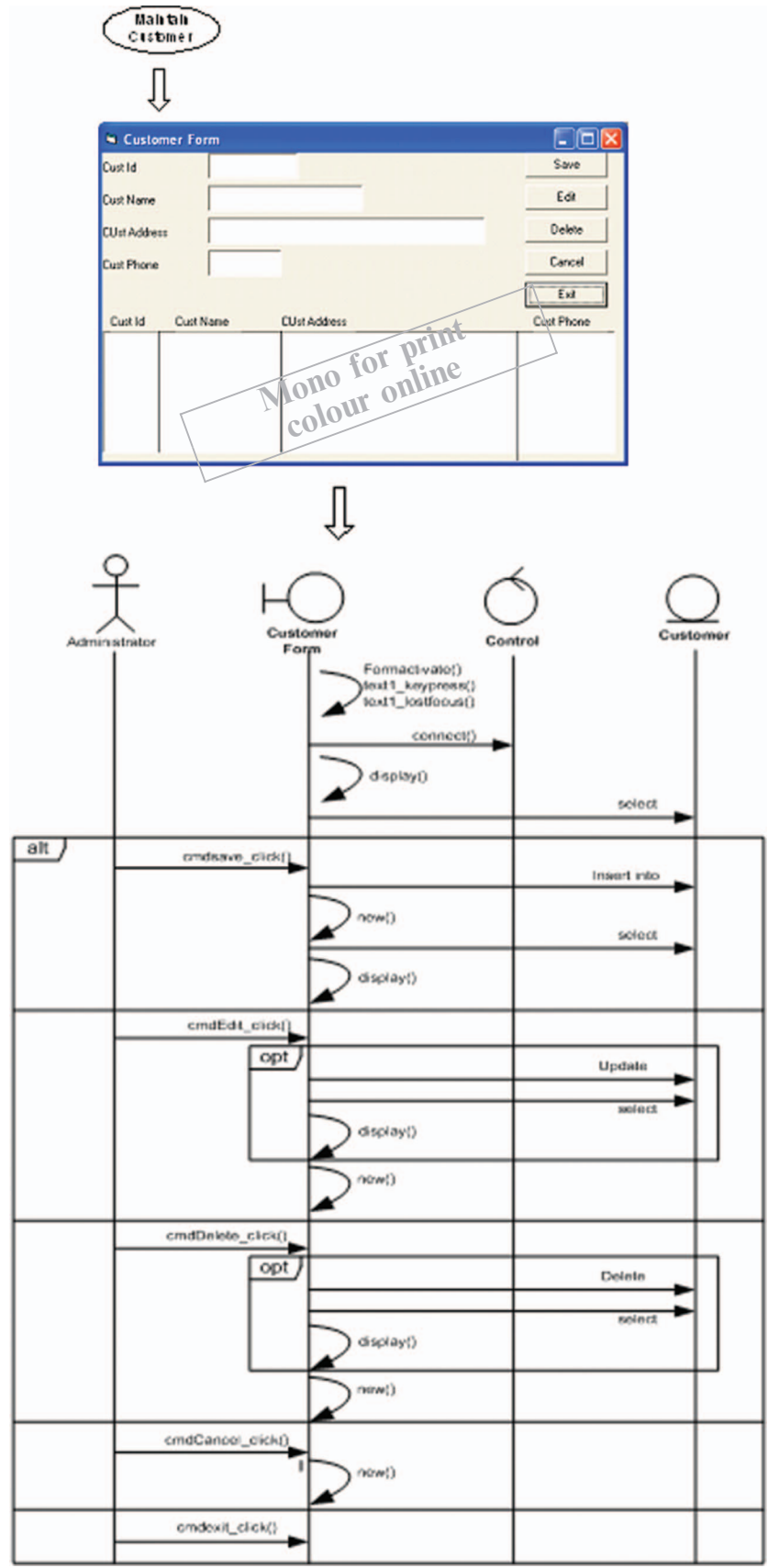

Figure 29. From use case to form and to sequence.

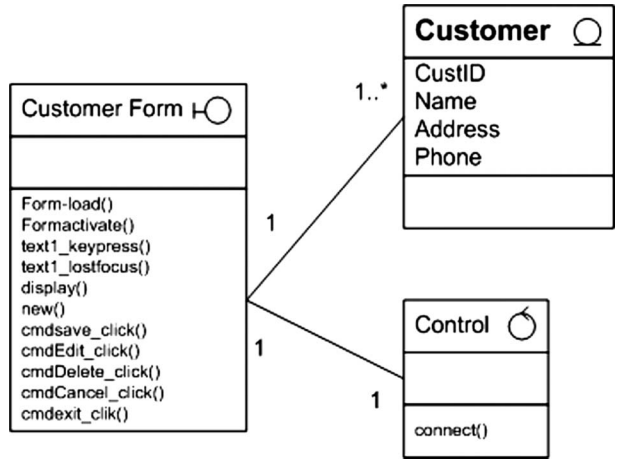

Figure 30.  Class relation on maintain customer sequence diagram.

where each of boundary class diagrams will have one form Visual Basic as prototyping implementation. Figure 28 shows the prototyping form with Visual Basic for all boundary class diagram.

Each use case diagram will have one boundary class and will have one form Visual Basic as prototyping implementation and has one sequence diagram. Then we will have seven sequence diagrams for seven use case on use case diagram and for seven boundary class diagrams. For example, we will use maintain customer use case, as shown in Figure 29.

In Customer form we have five command buttons then we have five alternatives as an optional on alt interaction frame. Entity class customer just only attach with sql statement and not method, because for the implementation will be used only with MS Access database. For control class name control have been invoked method connect( ), then automatically that method will become as method on control class name control. Also for customer form boundary class all the methods which have been invoked will become its methods. Figure 30 shows the relationship between boundary class diagram customer with entity class diagram customer and the relationship between boundary class diagram with control class diagram.

## 6.  Conclusion

Each use case on use case diagram can be attached with use case specification to explain what activity happens on that use case.

Other diagrams like activity diagram or state machine diagram can be used as extending diagrams to provide a picture about business process modelling in both as-is system or in to-be system. There is no need to use communication diagram as another way to picture interaction diagram beside sequence diagram.

Component diagram can be mixed with deployment diagram to picture for client–server implementation, client programming, server programming, two tier and three tier networking. All those diagrams and other new UML 2 diagram like composite structure diagram, timing diagram and interaction overview diagram can be used but keep it simple as agile methodology's requirement.

The next extended paper will be needed to explain the use of sequence diagram as dynamic diagram, besides the fact that each language program has its own syntax and characteristics.

CASE tool can be developed by this consistent approach both forward engineering and reverse engineering as an answer for Larman's issue about lack from CASE tools which can only do the static model (Larman 2005).

UML is a powerful tool to model the information system, one thing that we need is the person not having good object-oriented model and implementation programming skills will just draw bad UML design (Larman 2005). We need to make consistent our current knowledge between object-oriented language programming and object-oriented modelling. UML is a modelling language which has been approved by some industry leaders as a barometer to model object-oriented modelling. For some people UML is still lacking, but realistically UML is a current modelling language.